\documentclass[pra,twocolumn,showpacs,preprintnumbers,superscriptaddress]{revtex4}
\usepackage{times}
\usepackage{bm}
\usepackage{graphicx}
\usepackage{amsbsy}
\usepackage{amsmath}
\usepackage{amsfonts}
\usepackage{amsthm}

\begin{document}

\theoremstyle{plain}
\newtheorem{theorem}{Theorem}
\newtheorem{lemma}[theorem]{Lemma}
\newtheorem{corollary}[theorem]{Corollary}
\newtheorem{proposition}[theorem]{Proposition}
\newtheorem{conjecture}[theorem]{Conjecture}

\theoremstyle{definition}
\newtheorem{definition}[theorem]{Definition}

\theoremstyle{remark}
\newtheorem*{remark}{Remark}
\newtheorem{example}{Example}
\title{Probabilistic Teleportation of a Single Qubit: Unearthing New W-Class of States}
\author{Satyabrata Adhikari}
\email{satyabrata@dtu.ac.in} \affiliation{Delhi Technological
University, Shahbad Daulatpur, Main Bawana Road, Delhi-110042, Delhi, India}

\begin{abstract}
In this work, we propose a probabilistic teleportation protocol to teleport a single qubit via three-qubit W-states using two-qubit measurement basis. We show that for the proper choice of the state parameter of the resource state, it is possible to make success probability of the protocol very high. We deduce the condition for the successful execution of our teleportation protocol and this gives us new class of three-qubit W-states which act as a resource state. We have constructed operators that can be used to verify the condition of teleportation in experiment. This verification is necessary for the detection of whether the given three-qubit state is useful in our teleportation protocol or not. Further we quantify the amount of entanglement contained in the newly identified shared W-class of states. Moreover, we show that the W-class of shared state used in the teleportation protocol can be prepared using NMR set up.
\end{abstract}
\maketitle

\section{Introduction}
Entanglement, a quantum mechanical feature which has no analogue in classical domain \cite{horodecki4}. It act as a powerful resource in quantum information processing in the sense that without entanglement, a quantum state transfer or quantum teleportation of a quantum state would not be possible within the laws of quantum mechanics \cite{pirandola}. Quantum teleportation is an indispensable tool in quantum computing \cite{baur} and quantum communication \cite{gisin}. A revolutionary idea in the field of quantum communication had been given by Bennett et. al in 1993 in the form of a teleportation protocol \cite{bennett}. In their work, it has been shown that the information encoded in a single qubit can be transferred to a distant receiver via a two-qubit maximally entangled state shared between sender and receiver with the help of two bits of classical communication.
The protocols for quantum teleportation  are crucial not only in the development of quantum information theory but also in the enhancement of quantum technologies. The teleportation protocol can be deterministic \cite{bennett} or probabilistic \cite{li1} in the sense that the qubit can be teleported with unit probability or with some non-zero probability less than unity. Now if it is possible to teleport a qubit from one place to another distant place with unit fidelity and unit probability then it can be referred to as perfect teleportation. In the original teleportation protocol, perfect teleportation of a single qubit has been achieved with shared pure two-qubit maximally entangled state. However, if the shared entangled state is a non-maximally entangled state then there exist teleportation protocols by which we can teleport a qubit with unit fidelity but with some probability less than unity \cite{agrawal1}. There are some other teleportation protocol such as port-based teleportation protocol \cite{ishizaka} in which unknown quantum state is to be teleported to one of several ports at distant partner's site, symmetric multiparty-controlled teleportation protocol \cite{deng} where teleportation of an arbitrary two-qubit entangled state has been studied using two GHZ states \cite{greenberger}, and perfect controlled teleportation protocol \cite{li} where three-qubit entangled state as a shared resource state has been used. Neves et.al. have studied the teleportation of unknown qudit states through pure quantum channels with non-maximal Schmidt rank \cite{neves}. Therefore, the shared entangled state plays a vital role in the architecture of teleportation protocols. Experimental realization of quantum teleportation has also been successfully demonstrated with photonic qubits \cite{bouwmeester} and atomic qubits \cite{barrett}.\\
Two-qubit mixed entangled state can also be used as a resource in teleportation protocol to teleport a qubit \cite{verstraete}.  H. Jeong have studied quantum teleportation of optical qubits using hybrid entanglement as a quantum channel under decoherence effects \cite{jeong}. But it can be seen that touching the mark of unit fidelity is not possible in any teleportation protocol that involve mixed two-qubit entangled state as a shared resource state. Thus, shared mixed two-qubit entangled state cannot be considered as a good candidate for perfect teleportation.\\
It has been found that an unknown three-qubit GHZ-class of entangled state can be used as quantum channels in some teleportation scheme, which has been realized in experiment also \cite{shi,bouwmeester1}. Implementation of adiabatic quantum teleportation protocol for perfect teleportation with three-qubits has been investigated by Oh et.al \cite{oh}. It has also been found that a maximally entangled five-qubit state introduced by Brown et.al. \cite{brown} is useful for perfect teleportation of a single qubit and two-qubit state \cite{muralidharan}.\\
Apart from three-qubit GHZ class of states, there are another class of three-qubit states known as W-class of states. These two classes of three-qubit entangled states are inequivalent under stochastic local operation and classical communication (SLOCC) \cite{dur}.  We should note an important fact that if we trace out one qubit from three-qubit GHZ class of states then the reduced two-qubit state will become zero-discord state \cite{datta1} while the situation is different in case of three-qubit W-class of states. That is, if we trace out one qubit from three-qubit W-class of states then the reduced two-qubit mixed state would be an entangled state and not only that they may be used as a quantum channel in teleportation. In \cite{joo}, it has been shown that three-qubit W-state can serve as a useful shared resource state in quantum secure communication. Gorbachev et.al. \cite{gorbachev} have designed a quantum teleportation protocol using W-state as a shared quantum channel but this protocol demand nonlocal operation to recover the unknown state. There also exist other teleportation protocol, which uses W-state as a resource state but the protocol works with teleportation fidelity less than unity \cite{joo1}.\\
In 2006, Agrawal and Pati \cite{agrawal} have shown that there exist a three-qubit W-class of states that can be used as a shared resource state to achieve perfect teleportation of a single qubit state. Their teleportation protocol based on von Neumann type three-qubit projective measurement and two classical bit of communication. But in practical, it would be difficult to realize von Neumann type three-qubit projective measurement in the experiment. This motivate us to re-visit their teleportation protocol for any possibility to use two-qubit von Neumann type projective measurement instead of three-qubit projective measurement. In this work, we have shown that it is possible to resolve the issue noticed in Agrawal-Pati teleportation protocol by modifying their protocol in a way where we can use two-qubit projective measurement basis instead of three-qubit measurement basis. We overcome the problem by introducing probabilistic teleportation protocol in which it is possible to choose the state parameter of the resource state in such a way that the protocol will work with high probability of success. Hence, in this sense we will consider our protocol as nearly perfect teleportation protocol.\\
This paper is organized as follows: In Sec. II, we revisit Agrawal-Pati teleportation protocol to teleport a single qubit. In Sec. III, we introduce our teleportation protocol and derive the condition for which the single qubit can be teleported with unit fidelity but with probability near to unity. In Sec. IV, we discuss about the realization of our proposed nearly perfect teleportation protocol. In Sec. V, we quantify the amount of shared three-qubit entangled W-class of state used in our teleportation protocol and also discuss about its realization in experiment. In Sec. VI, we discuss the preparation of shared three-qubit W-states in the teleportation protocol using NMR set up. We conclude in Sec. VII.
\section{Revisiting Agrawal and Pati's perfect teleportation protocol to teleport a single qubit}
Agrawal and Pati studied a perfect teleportation protocol to teleport a single qubit, which followed original teleportation protocol with a difference that they have used three-qubit W-class of states as a shared resource state. Moreover, they have shown that there exist a special class of three-qubit W-state that can be used as a resource state for perfect teleportation of a single qubit. In this section, we revisit their teleportation scheme with some interesting facts.
\subsection{Agrawal and Pati's teleportation protocol}
Let us consider a single qubit state to be teleported as
\begin{eqnarray}
 |\psi\rangle_{x}=\alpha |0\rangle + \beta |1\rangle, |\alpha|^{2}+|\beta|^{2}=1
 \label{singlequbit}
\end{eqnarray}
Assume that the pure three qubit resource state belonging to W-class is given by
\begin{eqnarray}
|W(\lambda_{0},\lambda_{2},\lambda_{3})\rangle_{ABC}= \lambda_{0} |100\rangle + \lambda_{2} |001\rangle + \lambda_{3} |010\rangle
 \label{resourcequbit}
\end{eqnarray}
The normalization condition for the state $|W(\lambda_{0},\lambda_{2},\lambda_{3})\rangle_{ABC}$ is given by
\begin{eqnarray}
|\lambda_{0}|^{2}+|\lambda_{2}|^{2}+|\lambda_{3}|^{2}=1
 \label{normalcond}
\end{eqnarray}
Further, we assume that
the qubits 'x', 'A' and 'B' possessed by the sender Alice and the remaining qubit is with the receiver Bob.
The composite system of four qubits is described by the tensor product of $|\psi\rangle_{x}$ and $|W\rangle_{ABC}$ and it is given by
\begin{eqnarray}
 |\Phi\rangle_{xABC}&=&|\psi\rangle_{x}\otimes |W\rangle_{ABC}\nonumber\\&=&
 \alpha\lambda_{0} |0100\rangle+\alpha\lambda_{3} |0010\rangle+\alpha\lambda_{2} |0001\rangle+
 \nonumber\\&&\beta\lambda_{0} |1100\rangle+\beta\lambda_{3}|1010\rangle+\beta\lambda_{2} |1001\rangle
 \nonumber\\&=&\frac{1}{2}[|M_{1}^{+}\rangle_{xAB}\otimes(\alpha |0\rangle_{c} + \beta |1\rangle_{c})+
 |M_{1}^{-}\rangle_{xAB}\otimes \nonumber\\&& (\alpha |0\rangle_{c} - \beta |1\rangle_{c})+
 |M_{2}^{+}\rangle_{xAB}\otimes(\beta |0\rangle_{c} + \alpha |1\rangle_{c})\nonumber\\&&+
 |M_{2}^{-}\rangle_{xAB}\otimes(\beta |0\rangle_{c} - \alpha |1\rangle_{c})]
 \label{compositesystem}
 \end{eqnarray}
 where the three-qubit state vectors $|M_{1}^{+}\rangle_{xAB}$, $|M_{1}^{-}\rangle_{xAB}$, $|M_{2}^{+}\rangle_{xAB}$ and $|M_{2}^{-}\rangle_{xAB}$ are given by
 \begin{eqnarray}
 &&|M_{1}^{+}\rangle_{xAB}=\lambda_{0} |010\rangle+\lambda_{3} |001\rangle+\lambda_{2}|100\rangle\nonumber\\&&
 |M_{1}^{-}\rangle_{xAB}=\lambda_{0} |010\rangle+\lambda_{3} |001\rangle-\lambda_{2}|100\rangle)\nonumber\\&&
 |M_{2}^{+}\rangle_{xAB}=\lambda_{0} |110\rangle+\lambda_{3} |101\rangle+\lambda_{2}|000\rangle)\nonumber\\&&
 |M_{2}^{-}\rangle_{xAB}=\lambda_{0} |110\rangle+\lambda_{3} |101\rangle-\lambda_{2}|000\rangle)
 \label{mb}
 \end{eqnarray}
 The three-qubit state vectors $|M_{1}^{+}\rangle_{xAB}$, $|M_{1}^{-}\rangle_{xAB}$, $|M_{2}^{+}\rangle_{xAB}$ and $|M_{2}^{-}\rangle_{xAB}$ should be orthogonal to each other to realize the perfect teleportation but it can be easily shown that $\langle M_{1}^{+}|M_{1}^{-}\rangle \neq 0$ and $\langle M_{2}^{+}|M_{2}^{-}\rangle \neq 0$. Since all pairs of three-qubit state vectors are not orthogonal to each other so we have to impose orthogonality condition on the above three-qubit vectors.\\
The three-qubit vectors $|M_{1}^{+}\rangle_{xAB}$, $|M_{1}^{-}\rangle_{xAB}$, $|M_{2}^{+}\rangle_{xAB}$ and $|M_{2}^{-}\rangle_{xAB}$ are orthogonal
to each other if
\begin{eqnarray}
|\lambda_{0}|^2+|\lambda_{3}|^2=|\lambda_{2}|^2
\label{orthogonalitycond}
\end{eqnarray}
Under the condition given by (\ref{orthogonalitycond}), Alice can perform measurement on the state $|\Phi\rangle_{xABC}$ using three-qubit measurement basis $B=\{|M_{1}^{+}\rangle_{xAB}, |M_{1}^{-}\rangle_{xAB}, |M_{2}^{+}\rangle_{xAB} |M_{2}^{-}\rangle_{xAB}\}$. Depending on the measurement outcome, Alice send two bit of classical information to Bob and then Bob apply corresponding single qubit unitary operation to complete the teleportation protocol.
\subsection{Geometrical interpretation of the condition of perfect teleportation}
Using the orthogonality condition (\ref{orthogonalitycond}) in the normalization
condition (\ref{normalcond}), we have
\begin{eqnarray}
|\lambda_{2}|^2 = \frac{1}{2}
 \label{cond1}
\end{eqnarray}
From (\ref{orthogonalitycond}) and (\ref{cond1}), we have
\begin{eqnarray}
|\lambda_{0}|^2+|\lambda_{3}|^2=\frac{1}{2}
 \label{cond2}
\end{eqnarray}
Equation (\ref{cond2}) gives the required condition for perfect teleportation
of a single qubit using W-class state given by (\ref{resourcequbit}) as a resource state following
Agrawal-Pati's teleportation protocol.\\
Since $\lambda_{0}$ and $\lambda_{3}$ are complex parameters so we can always take $\lambda_{0}=ue^{i\theta_{1}}$ and $\lambda_{3}=ve^{i\theta_{2}}$, where $u$, $v$ are real variables and $\theta_{1}$, $\theta_{2}$ denote phases.\\
The equation (\ref{cond2}) then reduces to
\begin{eqnarray}
 u^2+v^2=\frac{1}{2}
\end{eqnarray}
Geometrically, it represent a circle with center at (0,0) and radius $\frac{1}{\sqrt{2}}$. At the center of the circle,
the parameters $\lambda_{0}$ and $\lambda_{3}$ are equal to zero and thus the state $|001\rangle$
lying at the center of the circle. It is interesting to note that the states lying on the circumference of the circle are useful for perfect teleportation in Agrawal-Pati's teleportation protocol.\\
Without any loss of generality, we can choose $u=\frac{1}{\sqrt{2+2n}}$ and $v=\frac{\sqrt{n}}{\sqrt{2+2n}}$, where $n$ is any real number.
Then the three-qubit W-class of states used as a resource state in Agrawal-Pati's teleportation protocol is given by
\begin{eqnarray}
&&|W(\frac{e^{i\theta_{1}}}{\sqrt{2+2n}},\frac{1}{\sqrt{2}},\frac{\sqrt{n}e^{i\theta_{2}}}{\sqrt{2+2n}})\rangle_{ABC}=
\frac{e^{i\theta_{1}}}{\sqrt{2+2n}}|100\rangle \nonumber\\&&+ \frac{1}{\sqrt{2}}|001\rangle + \frac{\sqrt{n}e^{i\theta_{2}}}{\sqrt{2+2n}}|010\rangle
 \label{resource3qubitstate}
\end{eqnarray}
If we take the parameter $n$ in such a way that either
\begin{eqnarray}
 u^2+v^2<\frac{1}{2}
\end{eqnarray}
or
\begin{eqnarray}
 u^2+v^2>\frac{1}{2}
\end{eqnarray}
holds then the three-qubit states lying inside or outside the circle and thus they are not useful as a resource state for perfect teleportation.
\section{Proposal of nearly perfect teleportation protocol to teleport a single qubit with pure three-qubit W-class resource state and two-qubit measurement basis}
In this section, we propose a teleportation protocol that require two-qubit Bell state measurement instead of three-qubit projective measurement to achieve nearly perfect teleportation of a single qubit. Moreover, we show that there exist another W-class of states different from the class introduced in \cite{agrawal} that can be used for nearly perfect teleportation of a single qubit.\\
 Let us start the teleportation protocol by assuming that the sender Alice want to send some information encoded in a single qubit state given in (\ref{singlequbit}) by teleporting the qubit to the receiver Bob. We consider again that Alice and Bob share the resource state (\ref{resourcequbit}). The resource state used in the protocol although belong to the class of W-state but we will show that they represent different class of W-state in comparison to the W-class given in (\ref{resource3qubitstate}).
\subsection{Proposal for nearly perfect teleportation protocol to teleport a single qubit}
Let us consider the composite system of four qubits which can be expressed as the tensor product of (i) a single qubit to be teleported given in (\ref{singlequbit}) and (ii) the three-qubit resource state given in (\ref{resourcequbit}). Therefore, four qubit state can be expressed as
\begin{eqnarray}
 |\Phi\rangle_{xABC}&=&|\psi\rangle_{x}\otimes |W\rangle_{ABC}\nonumber\\&=&
 \alpha\lambda_{0} |0100\rangle+\alpha\lambda_{3} |0010\rangle+\alpha\lambda_{2} |0001\rangle+
 \nonumber\\&&\beta\lambda_{0} |1100\rangle+\beta\lambda_{3}|1010\rangle+\beta\lambda_{2} |1001\rangle
 \nonumber\\&=&\frac{1}{2}[|N_{1}^{+}\rangle_{xAB}\otimes(\alpha |0\rangle_{c} + \beta |1\rangle_{c})+
 |N_{1}^{-}\rangle_{xAB}\otimes \nonumber\\&& (\alpha |0\rangle_{c} - \beta |1\rangle_{c})+
 |N_{2}^{+}\rangle_{xAB}\otimes(\beta |0\rangle_{c} + \alpha |1\rangle_{c})\nonumber\\&&+
 |N_{2}^{-}\rangle_{xAB}\otimes(\beta |0\rangle_{c} - \alpha |1\rangle_{c})]
\end{eqnarray}
 where the three-qubit vectors $|N_{1}^{+}\rangle_{xAB}$, $|N_{1}^{-}\rangle_{xAB}$, $|N_{2}^{+}\rangle_{xAB}$ and $|N_{2}^{-}\rangle_{xAB}$ can be expressed in the form as
 \begin{eqnarray}
 &&|N_{1}^{+}\rangle_{xAB}=(\lambda_{0} |01\rangle + \lambda_{2} |10\rangle) \otimes |0\rangle + \lambda_{3} |00\rangle \otimes |1\rangle \nonumber\\&&
 |N_{1}^{-}\rangle_{xAB}=(\lambda_{0} |01\rangle+\lambda_{2} |10\rangle) \otimes |0\rangle + \lambda_{3} |00\rangle \otimes |1\rangle \nonumber\\&&
 |N_{2}^{+}\rangle_{xAB}=(\lambda_{0} |11\rangle + \lambda_{2}|00\rangle) \otimes |0\rangle + \lambda_{3} |10\rangle \otimes |1\rangle \nonumber\\&&
 |N_{2}^{-}\rangle_{xAB}=(\lambda_{0} |11\rangle - \lambda_{2}|00\rangle) \otimes |0\rangle + \lambda_{3} |10\rangle \otimes |1\rangle
 \end{eqnarray}
Alice then construct the projectors $F$ and $I-F$, where $F=\frac{I_{xAB}+I_{xA} \otimes (\sigma_{z})_{B}}{2}$ for performing measurement on her qubits.\\
After the action of the projector $F$, the four qubit state $|\Phi\rangle_{xABC}$ can be expressed as
\begin{eqnarray}
 |\Upsilon^{(1)}\rangle_{xACB}&=&(F\otimes I_{C})|\Phi\rangle_{xABC}\nonumber\\&=& |\Psi^{(1)}\rangle_{xAC}\otimes |0\rangle_{B}
\end{eqnarray}
where the three-qubit state vector $|\Psi^{(1)}\rangle_{xAC}$ is given by
\begin{eqnarray}
|\Psi^{(1)}\rangle_{xAC}&=& \frac{1}{2}[|P_{1}^{+}\rangle_{xA}\otimes(\alpha |0\rangle_{c} + \beta |1\rangle_{c})+
 |P_{1}^{-}\rangle_{xA}\otimes \nonumber\\&& (\alpha |0\rangle_{c} - \beta |1\rangle_{c})+
 |P_{2}^{+}\rangle_{xA}\otimes(\beta |0\rangle_{c} + \alpha |1\rangle_{c})\nonumber\\&&+
 |P_{2}^{-}\rangle_{xA}\otimes(\beta |0\rangle_{c} - \alpha |1\rangle_{c})]
\end{eqnarray}
The vectors $|P_{1}^{+}\rangle_{xA}$, $|P_{1}^{-}\rangle_{xA}$, $|P_{2}^{+}\rangle_{xA}$,
$|P_{2}^{-}\rangle_{xA}$ are given by
\begin{eqnarray}
 &&|P_{1}^{+}\rangle_{xA}=\lambda_{0} |01\rangle + \lambda_{2} |10\rangle \nonumber\\&&
 |P_{1}^{-}\rangle_{xA}=\lambda_{0} |01\rangle - \lambda_{2} |10\rangle \nonumber\\&&
 |P_{2}^{+}\rangle_{xA}=\lambda_{0} |11\rangle + \lambda_{2}|00\rangle \nonumber\\&&
 |P_{2}^{-}\rangle_{xAB}=\lambda_{0} |11\rangle - \lambda_{2}|00\rangle
\end{eqnarray}
The probability that the state collapsing to $|\Psi^{(1)}\rangle_{xAC}\otimes |0\rangle_{B}$, after performing measurement with the projector $F$ is given by
\begin{eqnarray}
P^{(1)}=|\lambda_{0}|^{2}+|\lambda_{2}|^{2}
\label{prob1}
\end{eqnarray}
On the other hand, if Alice apply the projector $I-F$ on her qubit then the four qubit state $|\Phi\rangle_{xABC}$ reduces to
\begin{eqnarray}
 |\Upsilon^{(2)}\rangle_{xACB}&=&((I-F)\otimes I_{C})|\Phi\rangle_{xABC}\nonumber\\&=& |\Psi^{(2)}\rangle_{xAC}\otimes |1\rangle_{B}
\end{eqnarray}
where the three-qubit state vector $|\Psi^{(2)}\rangle_{xAC}$ is given by
\begin{eqnarray}
|\Psi^{(2)}\rangle_{xAC}&=& \lambda_{3}(\alpha|0\rangle+\beta|1\rangle)_{x}\otimes|0\rangle_{A}\otimes|0\rangle_{C}
\label{threequbitstatei-f}
\end{eqnarray}
The probability of getting the state $|\Psi^{(2)}\rangle_{xAC}\otimes |1\rangle_{B}$ is given by
\begin{eqnarray}
P^{(2)}=|\lambda_{3}|^{2}
\label{prob2}
\end{eqnarray}
It is clear from (\ref{threequbitstatei-f}) that if Alice perform projective measurement with the projector $I-F$ then the state cannot be teleported and hence our teleportation protocol fails.\\
To make our teleportation protocol nearly perfect, we can choose small value of the parameter $\lambda_{3}$ in the shared state so that the second
order term $\lambda_{3}^{2}$ is negligible. Therefore, it is possible to make the probability of getting the state $|\Psi^{(2)}\rangle_{xAC}\otimes |1\rangle_{B}$ almost zero. Alternatively, we can say that it is possible to take the probability $P_{1}$ toward unity and hence this make our protocol nearly perfect.\\
Let us now consider the case when Alice apply the projector $F$. In this scenerio, the two-qubit vectors $|P_{1}^{+}\rangle_{xA}$, $|P_{1}^{-}\rangle_{xA}$, $|P_{2}^{+}\rangle_{xA}$ and $|P_{2}^{-}\rangle_{xA}$ should be orthogonal to each other but we find that there exist at least one pair which are not orthogonal so we have to impose orthogonality condition on the above two-qubit vectors. The two-qubit vectors $|P_{1}^{+}\rangle_{xAB}$, $|P_{1}^{-}\rangle_{xAB}$, $|P_{2}^{+}\rangle_{xAB}$ and $|P_{2}^{-}\rangle_{xAB}$ are orthogonal to each other if
 \begin{eqnarray}
 |\lambda_{0}|^2=|\lambda_{2}|^2
 \label{orthogonalitycondnew}
\end{eqnarray}
Under the condition given by (\ref{orthogonalitycondnew}), Alice perform Bell-state measurement on her two qubits "x" and "A" and send the measurement outcome to Bob using two classical bits. In the last step of the protocol, Bob apply suitable Pauli operator on his qubit according to the Alice's measurement result to retrieve the original qubit given in (\ref{singlequbit}).\\

\subsection{Geometrical interpretation of the condition of nearly perfect teleportation protocol}
Using the orthogonality condition (\ref{orthogonalitycondnew}), the normalization condition (\ref{normalcond}) reduces to
\begin{eqnarray}
2|\lambda_{0}|^2+|\lambda_{3}|^2=1
\label{newcond}
\end{eqnarray}
where $|\lambda_{3}|$ is very small.\\
We therefore arrived at the condition (\ref{newcond}) which is the required condition for nearly perfect teleportation
of a single qubit using W-class state given by (\ref{resourcequbit}) as a resource state in our teleportation protocol.\\
The condition for nearly perfect teleportation given in (\ref{newcond}) can be re-expressed as
\begin{eqnarray}
\frac{|\lambda_{0}|^2}{(\frac{1}{\sqrt{2}})^{2}}+\frac{|\lambda_{3}|^2}{1^{2}}=1
\label{geomnewcond}
\end{eqnarray}
Choosing $\lambda_{0}=ue^{i\eta_{1}}$ and $\lambda_{3}=ve^{i\eta_{2}}$, equation (\ref{geomnewcond}) take the form
\begin{eqnarray}
\frac{u^{2}}{(\frac{1}{\sqrt{2}})^{2}}+\frac{v^{2}}{1^{2}}=1
\label{geomnewcond1}
\end{eqnarray}
where $u$, $v$ are real variables and $\eta_{1}$, $\eta_{2}$ are phases and $v$ is very small.\\
Geometrically, equation (\ref{geomnewcond1}) represents an ellipse with center at (0,0). The state $|001\rangle$
lying at the center of the ellipse. The length of the semi-major and semi-minor axis is $\frac{1}{\sqrt{2}}$ and
$1$ respectively. It can be observe that the three-qubit W-class of state used as a resource state in the modified perfect teleportation protocol are lying on the circumference of the ellipse.\\
If we choose the values of the real variables $u$ and $v$ in such a way that either
\begin{eqnarray}
\frac{u^{2}}{(\frac{1}{\sqrt{2}})^{2}}+\frac{v^{2}}{1^{2}}<1
\end{eqnarray}
or
\begin{eqnarray}
\frac{u^{2}}{(\frac{1}{\sqrt{2}})^{2}}+\frac{v^{2}}{1^{2}}>1
\end{eqnarray}
holds then the three-qubit states lying either inside or outside the ellipse and thus they are not useful as a resource state for perfect teleportation in modified teleportation protocol.
\subsection{New class of shared three-qubit W-states useful in teleportation of a single qubit}
In this section, we provide the actual form of a class of shared three-qubit W-states useful in modified teleportation protocol to teleport a single qubit.\\
Without any loss of generality, we can choose $u=\frac{\sqrt{m}}{\sqrt{2+2m}}$ and $v=\frac{\sqrt{2}}{\sqrt{2+2m}}$, where $m$ is any large real number.
Then the form of shared three-qubit W-class of states in modified teleportation protocol may be represented as
\begin{eqnarray}
&&|W_{s}(\sqrt{\frac{m}{2+2m}}e^{i\eta_{1}},\sqrt{\frac{m}{2+2m}},\sqrt{\frac{2}{2+2m}}e^{i\eta_{2}})\rangle_{ABC}\nonumber\\&=&
\frac{1}{\sqrt{2+2m}}[\sqrt{m}(e^{i\eta_{1}}|100\rangle + |001\rangle) + \sqrt{2}e^{i\eta_{2}}|010\rangle]
 \label{resource3qubitstate1}
\end{eqnarray}
\subsection{Comparison of our proposed teleportation protocol with Agrawal-Pati's protocol}
We are now in a position to compare modified teleportation protocol with Agrawal-Pati's teleportation protocol. The motivation of both the teleportation protocols is same, that is, both protocols work for teleporting a single qubit using shared W-class of state but there are significant differences between the two protocols. The differences will be discussed in the points given below:\\
(i) Agrawal-Pati teleportation protocol is deterministic but our proposed teleportation protocol is probabilistic with high probability of success.\\
(ii) Agrawal-Pati teleportation protocol needs three-qubit measurement for the successful implementation of the protocol while our protocol needs two-qubit Bell-state measurement when Alice perform the projective measurement $F$. Possibly, two-qubit measurements are easier to implement in an experiment than the three-qubit measurement. Further, we should note an important fact that discrimination of  four entangled Bell states is necessary for the implementation of teleportation protocol. In this context, we are fortunate since discrimination of  four entangled Bell states has been achieved in experiment with linear optical elements \cite{pavicic,houwelingen}.\\
(iii) We find that in our protocol, the shared W-class of states useful for perfect teleportation of a single qubit lies on an ellipse with center at (0,0) and length of semi-major and semi-minor axes of the ellipse are $\frac{1}{\sqrt{2}}$ and $1$ respectively but in Agrawal-Pati's teleportation protocol, shared W-class of states useful for perfect teleportation lies on the circumference of a circle with center at (0,0) and radius $\frac{1}{\sqrt{2}}$.\\
(iv) We find that the shared three-qubit W-class of states useful in our teleportation protocol lies on an ellipse and the perimeter of it is equal to $\sqrt{3}\pi$ while in the Agrawal-Pati teleportation protocol, the shared three-qubit W-class of states are lying on a circle with circumference $\sqrt{2}\pi$. Since in this case, the perimeter of the ellipse is greater than the circumference of a circle so we can conclude that our proposed teleportation protocol identifies more three-qubit W-class of states compare to Agrawal-Pati teleportation protocol.
\section{Realization of the condition of teleportation}
In this section, we express the condition of perfect teleportation for Agrawal-Pati's teleportation protocol and condition of nearly perfect teleportation of our teleportation protocol in terms of the concurrences \cite{wootters} of two-qubit reduced states. The two-qubit reduced states can be obtained by tracing out one qubit from three-qubit W-class states given in (\ref{resource3qubitstate}) and (\ref{resource3qubitstate1}) respectively. By doing this, we will show that the condition of perfect/nearly perfect teleportation for the above discussed teleportation protocol can be realized in experiment.\\
Let us consider the three-qubit state given in the canonical form as
\begin{eqnarray}
|\Omega\rangle &=& \lambda_{0}|000\rangle + \lambda_{1}e^{i\varphi}|100\rangle + \lambda_{2}|101\rangle
+ \lambda_{3}|110\rangle \nonumber\\&+& \lambda_{4}|111\rangle
\label{canonical}
\end{eqnarray}
where $\lambda_{i}'s$ are non-negative real numbers satisfying $\lambda_{0}^{2}+\lambda_{1}^{2}+\lambda_{2}^{2}+\lambda_{3}^{2}+\lambda_{4}^{2}=1$.\\
For the three-qubit state $|\Omega\rangle$, the invariants under local unitary transformations are given by \cite{torun}
\begin{eqnarray}
&&\lambda_{0}\lambda_{4}=\frac{\sqrt{\tau_{ABC}}}{2}\nonumber\\&&
\lambda_{0}\lambda_{2}=\frac{C_{AC}}{2}\nonumber\\&&
\lambda_{0}\lambda_{3}=\frac{C_{AB}}{2}\nonumber\\&&
|\lambda_{2}\lambda_{3}-e^{i\varphi}\lambda_{1}\lambda_{4}|=\frac{C_{BC}}{2}
\label{invariant}
\end{eqnarray}
Since we are interested in W-class of states and would like to keep all the concurrences of two-qubit state non-zero
so we choose $\lambda_{4}=0$. Further without any loss of generality we can assume $\lambda_{1}=0$. Thus these choices of the
parameters $\lambda_{1}$ and $\lambda_{4}$ reduce the canonical form of three-qubit state to the W-class of state given by
\begin{eqnarray}
|W_{s}(\lambda_{0},\lambda_{2},\lambda_{3})\rangle= \lambda_{0}|000\rangle + \lambda_{2}|101\rangle + \lambda_{3}|110\rangle
\label{specialw}
\end{eqnarray}
For the W-class of state $|W_{s}(\lambda_{0},\lambda_{2},\lambda_{3})\rangle$, the invariants given in (\ref{invariant}) also reduces to
\begin{eqnarray}
&&2\lambda_{0}\lambda_{2}=C_{AC}\nonumber\\&&
2\lambda_{0}\lambda_{3}=C_{AB}\nonumber\\&&
2\lambda_{2}\lambda_{3}=C_{BC}
\label{invariant1}
\end{eqnarray}
Solving equations (\ref{invariant1}), we can express the parameters $\lambda_{0}$ and $\lambda_{3}$ in terms of $\lambda_{2}$ which
is given by
\begin{eqnarray}
\lambda_{0}=\frac{C_{AB}}{C_{BC}}\lambda_{2},~~ \lambda_{3}=\frac{C_{AB}}{C_{AC}}\lambda_{2}
\label{parameters}
\end{eqnarray}
\subsection{Case of Agrawal-Pati teleportation protocol}
Tajima studied the necessary and sufficient condition of the possible application of a deterministic LOCC transformation of three-qubit pure states \cite{tajima}. LOCC operation $\sigma_{x} \otimes I \otimes I$ reduces the state $|W_{s}(\lambda_{0},\lambda_{2},\lambda_{3})\rangle$ given by (\ref{specialw}) to the state $|W(\lambda_{0},\lambda_{2},\lambda_{3})\rangle_{ABC}$ given by (\ref{resourcequbit}). This implies that
\begin{eqnarray}
|W(\lambda_{0},\lambda_{2},\lambda_{3})\rangle_{ABC}&=&(\sigma_{x} \otimes I \otimes I)|W_{s}(\lambda_{0},\lambda_{2},\lambda_{3})\rangle_{ABC}\nonumber\\&=&\lambda_{0} |100\rangle + \lambda_{2} |001\rangle
+ \lambda_{3} |010\rangle
\label{transformation}
\end{eqnarray}
For Agrawal-Pati teleportation protocol, $\lambda_{2}=\frac{1}{\sqrt{2}}$. Thus, the parameters $\lambda_{0}$ and $\lambda_{3}$
given in (\ref{parameters}) can be re-expressed as
\begin{eqnarray}
\lambda_{0}=\frac{C_{AB}}{\sqrt{2}C_{BC}},~~ \lambda_{3}=\frac{C_{AB}}{\sqrt{2}C_{AC}}
\label{parameters1}
\end{eqnarray}
In this case, the condition for perfect teleportation given by (\ref{cond2}) can be re-written
in terms of concurrences $C_{AB}$, $C_{BC}$, $C_{AC}$ as
\begin{eqnarray}
\frac{1}{C_{AB}^{2}}=\frac{1}{C_{BC}^{2}}+\frac{1}{C_{AC}^{2}}
\label{apcondconc}
\end{eqnarray}
The relation (\ref{apcondconc}) can be expressed as
\begin{eqnarray}
C_{AB}^{2}=\frac{1}{2}H(C_{BC}^{2},C_{AC}^{2})
\label{apcondconc1}
\end{eqnarray}
where $H(C_{BC}^{2},C_{AC}^{2})=\frac{1}{C_{BC}^{2}}+\frac{1}{C_{AC}^{2}}$ denote the harmonic mean of $C_{BC}^{2}$ and $C_{AC}^{2}$.\\
Also, the relation between harmonic mean and geometric mean is given by
\begin{eqnarray}
H(C_{BC}^{2},C_{AC}^{2})\leq G(C_{BC}^{2},C_{AC}^{2})
\label{hmgm}
\end{eqnarray}
where $G(C_{BC}^{2},C_{AC}^{2})=C_{BC}~C_{AC}$ denote the geometric mean.\\
From (\ref{apcondconc1}) and (\ref{hmgm}), we have
\begin{eqnarray}
&&C_{AB}^{2}\leq \frac{1}{2}G(C_{BC}^{2},C_{AC}^{2})\nonumber\\&&
\Rightarrow C_{AB}^{2}\leq \frac{1}{2} C_{BC}C_{AC}
\label{apcondconc2}
\end{eqnarray}
The equality condition holds when $C_{BC}=C_{AC}$. Therefore, the inequality (\ref{apcondconc2}) reduces to
an equality and it is given by
\begin{eqnarray}
C_{AB}^{2}= \frac{1}{2} C_{AC}^{2}= \frac{1}{2} C_{BC}^{2}
\label{equrel}
\end{eqnarray}
Under the particular condition (\ref{equrel}), the parameters $\lambda_{0}$, $\lambda_{2}$, and $\lambda_{3}$ are
given by
\begin{eqnarray}
\lambda_{0}=\frac{1}{2},~~\lambda_{2}=\frac{1}{\sqrt{2}},~~ \lambda_{3}=\frac{1}{2}
\label{parameters2}
\end{eqnarray}
Therefore, the corresponding shared three-qubit W-class of states useful in Agrawal-Pati's teleportation protocol is given by
\begin{eqnarray}
|W(\frac{1}{2},\frac{1}{\sqrt{2}},\frac{1}{2})\rangle_{ABC}=\frac{1}{2} |100\rangle + \frac{1}{\sqrt{2}} |001\rangle
+ \frac{1}{2} |010\rangle
\label{particularstate}
\end{eqnarray}

\subsection{Case of our proposed teleportation protocol}
The condition of nearly perfect teleportation for our proposed teleportation protocol is given by (\ref{orthogonalitycondnew})
and it can be re-expressed in terms of the concurrences of the two-qubit reduced states as
\begin{eqnarray}
C_{AB}=C_{BC}
\label{scondconc}
\end{eqnarray}
The three-qubit W-class shared state useful for perfect teleportation for the teleportation of a
single qubit using our proposed teleportation protocol is
given by $|W_{s}(\lambda_{0},\lambda_{2},\lambda_{3})\rangle$, where the state parameters can be expressed in terms
of the concurrences $C_{AB}$ and $C_{AC}$ as
\begin{eqnarray}
&&\lambda_{0}^{2}=\lambda_{2}^{2}=\frac{C_{AC}^{2}}{2C_{AC}^{2}+C_{AB}^{2}}\nonumber\\&&
\lambda_{3}^{2}=\frac{C_{AB}^{2}}{2C_{AC}^{2}+C_{AB}^{2}}
\label{stateparametersnew}
\end{eqnarray}
We should note that for nearly perfect teleportation, the state parameter $\lambda_{3}$ is a small quantity and
hence the concurrence $C_{AB}$ must be very small.
\subsection{Discussion}
In \cite{datta}, it has been shown that there exist operators $O_{1}$,
$O_{2}$ and $O_{3}$ decomposed in terms of Pauli matrices, which can be
used to classify three-qubit pure states. The experimental classification of three-qubit pure states
based on these operators has been achieved \cite{singh}. The operators $O_{1}$,
$O_{2}$ and $O_{3}$ can be defined as \cite{datta}
\begin{eqnarray}
&&O_{1}= 2(\sigma_{x}\otimes \sigma_{x} \otimes \sigma_{z}),\nonumber\\&&
O_{2}= 2(\sigma_{x}\otimes \sigma_{z} \otimes \sigma_{x})\nonumber\\&&
O_{2}= 2(\sigma_{z}\otimes \sigma_{x} \otimes \sigma_{x})
\label{operators}
\end{eqnarray}
The square of the expectation values of the the above operators with respect to the
state $|W_{s}(\lambda_{0},\lambda_{2},\lambda_{3})\rangle$ are given by
\begin{eqnarray}
&&C_{AB}^{2}=\frac{\langle O_{1}\rangle_{W_{s}}^{2}}{4},\nonumber\\&&
C_{AC}^{2}=\frac{\langle O_{2}\rangle_{W_{s}}^{2}}{4}\nonumber\\&&
C_{BC}^{2}=\frac{\langle O_{3}\rangle_{W_{s}}^{2}}{4}
\label{expectationvalue}
\end{eqnarray}
Therefore, the concurrences $C_{AB}$, $C_{BC}$ and $C_{CA}$  of the two-qubit reduced states of three-qubit $W_{s}$-class of states
can be realized experimentally.\\
Thus, the condition of perfect teleportation given in (\ref{apcondconc}) for Agrawal-Pati teleportation protocol and the condition given in (\ref{scondconc}) for our proposed teleportation protocol can be realized experimentally. Hence, for any given shared W-class of states, we can easily verify experimentally whether the given state can be used for perfect/nearly perfect teleportation of a single qubit either by applying
Agrawal-Pati teleportation protocol or by our proposed teleportation protocol.
\section{Quantification of three-qubit entanglement using three-$\pi$ entanglement measure}
In this section, we would like to quantify the amount of entanglement in three-qubit W-class of states useful in our proposed teleportation protocol
by three-$\pi$ entanglement measure. An entanglement measure used to quantify three-qubit entanglement in terms of negativity referred to as three-${\pi}$ entanglement measure \cite{ou}. It can be defined as
\begin{eqnarray}
\pi_{ABC}=\frac{\pi_{A}+\pi_{B}+\pi_{C}}{3}
\label{three-pi}
\end{eqnarray}
where $\pi_{A}$, $\pi_{B}$, $\pi_{C}$ denote the residual entanglement given by
\begin{eqnarray}
\pi_{A}=N_{A(BC)}^{2}-N_{AB}^{2}-N_{AC}^{2}\nonumber\\
\pi_{B}=N_{B(CA)}^{2}-N_{BC}^{2}-N_{BA}^{2}\nonumber\\
\pi_{C}=N_{C(AB)}^{2}-N_{CA}^{2}-N_{CB}^{2}
\label{residualent}
\end{eqnarray}
For any pure three-qubit state, it has been shown that \cite{ou}
\begin{eqnarray}
&&N_{A(BC)}=C_{A(BC)}, N_{B(CA)}=C_{B(CA)},\nonumber\\&&
N_{C(AB)}=C_{C(AB)}
\label{negconcrel}
\end{eqnarray}
Therefore, the relations for residual entanglement given in (\ref{residualent}) reduces to
\begin{eqnarray}
\pi_{A}=C_{A(BC)}^{2}-N_{AB}^{2}-N_{AC}^{2}\nonumber\\
\pi_{B}=C_{B(CA)}^{2}-N_{BC}^{2}-N_{BA}^{2}\nonumber\\
\pi_{C}=C_{C(AB)}^{2}-N_{CA}^{2}-N_{CB}^{2}
\label{residualent1}
\end{eqnarray}
Now our task is to show that the three-$\pi$ entanglement measure, which measures the amount of entanglement
of three-qubit W-class states useful in the proposed teleportation protocol, can be realized experimentally. To move towards
our goal, we will propose the method of realizing the quantities $C_{A(BC)}^{2}$, $C_{B(CA)}^{2}$, $C_{C(AB)}^{2}$,
$N_{AB}^{2}$, $N_{AC}^{2}$ and $N_{CA}^{2}$ experimentally.
\subsection{Realization of $C_{A(BC)}^{2}$, $C_{B(CA)}^{2}$ and $C_{C(AB)}^{2}$}
The tangle $\tau_{ABC}$ for three qubits A, B, C can be defined by the relation
\begin{eqnarray}
C_{C(BA)}^{2}= \tau_{ABC}+C_{CA}^{2}+C_{CB}^{2}
\label{relation1}
\end{eqnarray}
where $C_{C(AB)}$ represent the concurrence between the qubit C and the pair of qubits B,A taking together, and $C_{AB}$, $C_{CA}$, $C_{CB}$ denote the concurrences of two qubit reduced states $\rho_{AB}$, $\rho_{BC}$, $\rho_{AC}$.\\
For W-class of states, $\tau_{ABC}=0$ and thus the relation (\ref{relation1}) reduces to
\begin{eqnarray}
C_{C(BA)}^{2}= C_{AC}^{2} + C_{BC}^{2}
\label{relation20}
\end{eqnarray}
Similar relations can be written for $C_{B(CA)}^{2}$, $C_{A(BC)}^{2}$ as
\begin{eqnarray}
C_{B(CA)}^{2}= C_{BC}^{2} + C_{BA}^{2}
\label{relation21}
\end{eqnarray}
\begin{eqnarray}
C_{A(BC)}^{2}= C_{AB}^{2} + C_{AC}^{2}
\label{relation22}
\end{eqnarray}
For the three-qubit W-class of states represented by $|W_{s}(\lambda_{0},\lambda_{2},\lambda_{3})\rangle$, the
relations (\ref{relation20}), (\ref{relation21}) and (\ref{relation22}) can be re-expressed in terms of the expectation
values of the operators $O_{1}$, $O_{2}$, $O_{3}$ as
\begin{eqnarray}
&&C_{A(BC)}^{2}= \frac{1}{4}(\langle O_{1}\rangle_{W_{s}}^{2}+\langle O_{2}\rangle_{W_{s}}^{2})\nonumber\\&&
C_{B(CA)}^{2}= \frac{1}{4}(\langle O_{1}\rangle_{W_{s}}^{2}+\langle O_{3}\rangle_{W_{s}}^{2})\nonumber\\&&
C_{C(AB)}^{2}= \frac{1}{4}(\langle O_{2}\rangle_{W_{s}}^{2}+\langle O_{3}\rangle_{W_{s}}^{2})\nonumber\\&&
\label{relation31}
\end{eqnarray}
Thus, the concurrences $C_{A(BC)}^{2}$, $C_{B(CA)}^{2}$ and $C_{C(AB)}^{2}$ can be realized experimentally.
\subsection{Realization of negativities for two-qubit reduced states}
Let us consider again the three-qubit $|W_{s}\rangle_{ABC}$ state as given in (\ref{specialw}) to find out the negativities of the two-qubit
reduced states described by the density matrices $\rho_{AB}=Tr_{C}(|W_{s}\rangle_{ABC}\langle W_{s}|$, $\rho_{BC}=Tr_{A}(|W_{s}\rangle_{ABC}\langle W_{s}|$, $\rho_{CA}=Tr_{B}(|W_{s}\rangle_{ABC}\langle W_{s}|$. Therefore, the negativities of the two-qubit reduced states $\rho_{AB}$, $\rho_{BC}$ and $\rho_{CA}$ are given by
\begin{eqnarray}
&&N_{AB}= \sqrt{\lambda_{2}^{4}+4\lambda_{0}^{2}\lambda_{3}^{2}}-\lambda_{2}^{2}\nonumber\\&&
N_{BC}= \sqrt{\lambda_{0}^{4}+4\lambda_{2}^{2}\lambda_{3}^{2}}-\lambda_{0}^{2}\nonumber\\&&
N_{CA}= \sqrt{\lambda_{3}^{4}+4\lambda_{2}^{2}\lambda_{0}^{2}}-\lambda_{3}^{2}
\label{negativity}
\end{eqnarray}
The three-qubit class of states useful for nearly perfect teleportation of a single qubit according to our proposed teleportation protocol is given by (\ref{specialw}) with the state parameters given by (\ref{stateparametersnew}).
Using equation (\ref{stateparametersnew}) in (\ref{negativity}), the negativities of the two-qubit reduced states can be expressed
as
\begin{eqnarray}
&&N_{AB}= N_{BC}= \frac{C_{AC}(\sqrt{C_{AC}^{2}+4C_{AB}^{2}}-C_{AC})}{2C_{AC}^{2}+C_{AB}^{2}}\nonumber\\&&
N_{CA}= \frac{\sqrt{C_{AB}^{4}+4C_{AC}^{4}}-C_{AB}^{2}}{2C_{AC}^{2}+C_{AB}^{2}}
\label{negativity1}
\end{eqnarray}
Equation (\ref{expectationvalue}) tells us that the negativities $N_{AB}$, $N_{BC}$, $N_{CA}$ given in (\ref{negativity1}) can also be realized experimentally.
\subsection{Realization of three-$\pi$ entanglement measure}
We are now in a position to use three-$\pi$ entanglement measure to measure the amount of entanglement in $W_{s}-$ class of states useful
for the teleportation of a single qubit following modified teleportation protocol.\\
The three-$\pi$ entanglement measure for the quantification of $W_{s}-$ class of states used in our teleportation protocol is given by
\begin{eqnarray}
\pi_{ABC}=4(C_{AB}^{2}-N_{AB}^{2})+ 2(C_{AC}^{2}-N_{AC}^{2})
\label{three-pinew}
\end{eqnarray}
where $N_{AB}$ and $N_{AC}$ are given by (\ref{negativity1}).\\
Since the two-qubit entanglement measure $N_{AB}$ and $N_{AC}$ can be expressed in terms of concurrences
$C_{AB}$ and $C_{AC}$ and as shown in (\ref{expectationvalue}) these entanglement measures can be realized
experimentally so three-$\pi$ entanglement measure $\pi_{ABC}$ can also be realized experimentally.
\section{Proposal to realize the condition of perfect teleportation in NMR set up}
A general three-qubit pure state is given by
\begin{eqnarray}
|\Theta\rangle_{ABC}&=&\textrm{cos}\alpha |000\rangle + \textrm{sin}\alpha~\textrm{cos}\beta~\textrm{sin}\gamma |001\rangle \nonumber\\&+& \textrm{sin}\alpha~\textrm{sin}\beta |010\rangle + \textrm{sin}\alpha~\textrm{cos}\beta~\textrm{cos}\gamma~\textrm{cos}\delta |100\rangle
\nonumber\\&+& e^{i\phi}\textrm{sin}\alpha~\textrm{cos}\beta~\textrm{cos}\gamma~\textrm{sin}\delta|111\rangle
\label{threequbit}
\end{eqnarray}
where the four parameters $\alpha \in [0,\frac{\pi}{2}]$, $\beta \in [0,\frac{\pi}{2}]$, $\gamma \in [0,\frac{\pi}{2}]$, $\delta \in [0,\frac{\pi}{2}]$ whereas the relative phase parameter $\delta$ lying within $0$ and $2\pi$ (including $0$ and $2\pi$).\\
The above general three-qubit pure state state can be constructed in NMR experiment by implementing a single qubit rotation gate, several two-qubit controlled-rotation and controlled-NOT gates, a three-qubit Toffoli gate and a controlled-controlled phase gate \cite{dogra}.\\
Choosing $\alpha=\frac{\pi}{2}$ and $\delta=\phi=0$, a canonical form of general three-qubit state (\ref{threequbit}) reduces to W-class state and it is given by
\begin{eqnarray}
|W\rangle_{ABC}&=&\textrm{cos}\beta~\textrm{sin}\gamma |001\rangle + \textrm{sin}\beta |010\rangle\nonumber\\&+& \textrm{cos}\beta~\textrm{cos}\gamma |100\rangle
\label{wclass}
\end{eqnarray}

\subsection{Preparation of shared entangled state for Agrawal-Pati teleportation protocol in NMR set up}
The preparation of shared entangled state is a necessary task to realize any teleportation protocol. Three-qubit W-class of
state has been used in Agrawal-Pati teleportation protocol and it can be prepared in NMR set up by choosing the appropriate value of the
parameter $\beta$ and $\gamma$.\\
The three-qubit W-class of state given in (\ref{wclass}) can be used as a resource state in Agrawal-Pati teleportation protocol if
\begin{eqnarray}
\textrm{cos}\beta~\textrm{sin}\gamma=\frac{1}{\sqrt{2}}
\label{condition}
\end{eqnarray}
Using the condition (\ref{condition}), we can construct W-class of states using NMR set up, which can be used as shared resource state in Agrawal-Pati teleportation protocol and the prepared class of states take the form as
\begin{eqnarray}
|W^{AP}\rangle_{ABC}&=&\frac{\sqrt{2\textrm{cos}^{2}\beta-1}}{\sqrt{2}} |100\rangle + \frac{1}{\sqrt{2}}|001\rangle\nonumber\\&+& \textrm{sin}\beta |010\rangle,~~~~\beta \in (0,\frac{\pi}{4}]
\label{expform}
\end{eqnarray}

\subsection{Preparation of shared entangled state for our teleportation protocol in NMR set up}
In our teleportation protocol, we have used different class of three-qubit W-state and it can be observe that this class of states can also be prepared in NMR set up.\\
The three-qubit W-class of state given in (\ref{resource3qubitstate1}) can be used as a resource state in our teleportation protocol if
\begin{eqnarray}
\textrm{tan}\gamma = 1,~~ i.e.~~ \gamma=\frac{\pi}{4}
\label{condition1}
\end{eqnarray}
Therefore, three-qubit W-class of states used in our teleportation protocol can be prepared using NMR set up by choosing the paramater $\gamma=\frac{\pi}{4}$. Hence, the form of the W-class can be expressed as
\begin{eqnarray}
|W^{M}\rangle_{ABC}&=&\frac{\textrm{cos}\beta}{\sqrt{2}} |100\rangle + \frac{\textrm{cos}\beta}{\sqrt{2}} |001\rangle\nonumber\\&+& \textrm{sin}\beta |010\rangle,~~~~\beta \in (0,\frac{\pi}{2})
\label{expform1}
\end{eqnarray}
For nearly perfect teleportation, the value of $\beta$ must be chosen in the neighbourhood of zero.\\
By noticing the range of the parameter $\beta$ appeared in two equations (\ref{expform}) and (\ref{expform1}), we can again
conclude that the set containing three-qubit W-class of states used in our teleportation protocol is larger than the set
of three-qubit W-class of states used in Agrawal-Pati teleportation protocol.
\section{Conclusion}
To summarize, we have discussed a teleportation scheme for achieving the nearly perfect teleportation of a single qubit using three-qubit W-class
of states as a shared quantum state. We found that the shared three-qubit W-states used in the proposed teleportation scheme belong to
different class of W-states than the class of W-states introduced by Agrawal and Pati. We have shown that the shared three-qubit states used in our teleportation protocol lies on an ellipse with center at (0,0). The identified new class of three-qubit W-states contains larger number of states in comparison to the number of three-qubit W-states contained in the class introduced by Agrawal and Pati. In our teleportation protocol, we need two-qubit measurement basis while Agrawal-Pati's teleportation protocol need three-qubit measurement basis. We have expressed the condition of the our teleportation scheme and Agrawal-Pati teleportation protocol in terms of the concurrences of the reduced two-qubit state obtained after tracing out one qubit from the three-qubit W-class of states. Since it has been shown in this work that the values of the concurrences of the reduced two-qubit state can be obtained experimentally so the conditions of teleportation can be verified experimentally. This verification will emphasize on the fact that whether the given three-qubit state could be used in our teleportation protocol or Agrawal-Pati teleportation protocol or not useful for teleportation. We have also shown that three-$\pi$ entanglement measure can be used to quantify the amount of entanglement of shared three-qubit W-class of states useful in the proposed teleportation protocol and also discussed its implementation in experiment. Lastly, we discuss the preparation of three-qubit W-class of shared states for the two teleportation protocol in NMR experiment.


\begin{thebibliography}{90}
\bibitem{horodecki4} R. Horodecki, P. Horodecki, M. Horodecki, and K. Horodecki, Rev. Mod. Phys. \textbf{81}, 865 (2009).
\bibitem{pirandola} S. Pirandola, J. Eisert, C. Weedbrook, A. Furusawa, and S. L. Braunstein, Nat. Photon. \textbf{9}, 641 (2015).
\bibitem{baur} M. Baur, A. Fedorov, L. Steffen, S. Filipp, M.P. da Silva, and A. Wallraff, Phys. Rev. Lett.
\textbf{108}, 040502 (2012).
\bibitem{gisin} N. Gisin, and R. Thew, Nat. phot. \textbf{ 1}, 165 (2007).
\bibitem{bennett} C. H. Bennett, G. Brassard, C. Crepeau, R. Jozsa, A. Peres, and W. K. Wootters, Phys. Rev. Lett.
\textbf{70}, 1895 (1993).
\bibitem{li1} W. L. Li, C. F. Li, and G. C. Guo, Phys. Rev. A \textbf{61}, 034301 (2000).
\bibitem{agrawal1} P. Agrawal and A. K. Pati, Phys. Lett. A \textbf{305}, 12 (2002).
\bibitem{ishizaka} S. Ishizaka and T. Hiroshima, Phys. Rev. Lett. \textbf{101}, 240501 (2008); D. P. Garcia, Phys. Rev. A \textbf{87}, 040303(R)
(2013).
\bibitem{deng} F-G Deng, C-Y Li, Y-S Li, H-Y Zhou, and Y. Wang, Phys. Rev. A \textbf{72}, 022338 (2005).
\bibitem{greenberger} M. Greenberger, M. A. Horne, A. Shimony and A. Zeilinger, Am. J. Phys \textbf{58}, 1131 (1990).
\bibitem{li} X-H Li, and S. Ghose, Phys. Rev. A \textbf{90}, 052305 (2014).
\bibitem{neves} L. Neves, M. A. Solis-Prosser, A. Delgado, and O. Jimenez, Phys. Rev. A \textbf{85}, 062322 (2012).
\bibitem{bouwmeester} D. Bouwmeester, J. W. Pan, K. Mattle, M. Eibl, H. Weinfurter, A. Zeilinger, Nature \textbf{390}, 575 (1997).
\bibitem{barrett} M. D. Barrett, J. Chiaverini, T. Schaetz, J. Britton, W. M. Itano, J. D. Jost, E. Knill, C. Langer, D. Leibfried, R. Ozeri, and D.
    J. Wineland, Nature \textbf{429}, 737 (2004).
\bibitem{verstraete} F. Verstraete, and H. Verschelde, Phys. Rev. Lett. \textbf{90} 097901 (2003); M. L. Hu, Eur. Phys. J. D \textbf{64}, 531–538 (2011).
\bibitem{jeong} H. Jeong, S. Bae, S. Choi, Quant. Inf. Proc. \textbf{15}, 913 (2016).
\bibitem{shi} B. S. Shi, Y. K. Jiang, G. C. Guo, Phys. Lett. A \textbf{268}, 161 (2000).
\bibitem{bouwmeester1} D. Bouwmeester, J. W. Pan, M. Daniell, H. Weinfurter, and A. Zeilinger, Phys. Rev. Lett. \textbf{82}, 1345 (1999).
\bibitem{oh} S. Oh, Y-P Shim, J. Fei, M. Friesen and X. Hu, Phys. Rev. A \textbf{87}, 022332 (2013).
\bibitem{brown} I. D. K. Brown, S. Stepney, A. Sudbery, and S. L. Braunstein, J. Phys. A \textbf{38}, 1119 (2005).
\bibitem{muralidharan} S. Muralidharan and P. K. Panigrahi, Phys. Rev. A \textbf{77}, 032321 (2008).
\bibitem{dur} W. Dur, G. Vidal, and J. I. Cirac, Phys. Rev. A \textbf{62}, 062314 (2000).
\bibitem{datta1} A. Datta, A. Shaji, C. M. Caves, Phys. Rev. Lett. \textbf{100}, 050502 (2008).
\bibitem{joo} J. Joo, J. Lee, J. Jang, and Y-J. Park, e-print quant-ph/0204003.
\bibitem{gorbachev} V. N. Gorbachev, A. A. Rodichkina, and A. I. Trubilko, Phys. Lett. A \textbf{310}, 339 (2003).
\bibitem{joo1} J. Joo, Y-J. Park, S. Oh, and J. Kim, New J. Phys. \textbf{5}, 136 (2003).
\bibitem{agrawal} P. Agrawal, and A. Pati, Phys. Rev. A \textbf{74}, 062320 (2006).
\bibitem{pavicic} M. Pavicic, Phys. Rev. Lett. \textbf{107}, 080403 (2011).
\bibitem{houwelingen} J. A. W. van Houwelingen, N. Brunner, A. Beveratos, H. Zbinden, and N. Gisin, Phys. Rev. Lett. \textbf{96}, 130502 (2006);
J. A. W. van Houwelingen, A. Beveratos, N. Brunner, N. Gisin, and H. Zbinden, Phys. Rev. A \textbf{74}, 022303 (2006).
\bibitem{wootters} W. K. Wootters, Phys. Rev. Lett. \textbf{80}, 2245 (1998).
\bibitem{torun} G. Torun, and A. Yildiz, Phys. Rev. A \textbf{89}, 032320 (2014).
\bibitem{tajima} H. Tajima, Annals of Phys. \textbf{329}, 1 (2013).
\bibitem{datta} C. Datta, S. Adhikari, A. Das and P. Agrawal, Eur. Phys. J. D \textbf{72}, 157 (2018).
\bibitem{singh} A. Singh, H. Singh, K. Dorai and Arvind, Phys. Rev. A \textbf{98}, 032301 (2018).
\bibitem{ou} Y. C. Ou, and H. Fan, Phys. Rev. A \textbf{75}, 062308 (2007).
\bibitem{dogra} S. Dogra, K. Dorai and Arvind, Phys. Rev. A \textbf{91}, 022312 (2015).
\end{thebibliography}
\end{document}